\documentstyle[preprint,aps]{revtex}
\begin{document}
\draft
\title{Electron capture on iron group nuclei}
\author{D.J. Dean$^{1,2}$, K. Langanke $^3$, L. Chatterjee$^{1,2,4}$,  
P.B. Radha$^{5}$, and M. R. Strayer$^1$}
\address{$^1$ Physics Division, Oak Ridge National Laboratory, P.O. Box 2008\\
Oak Ridge, Tennessee 37831-6373 USA \\
$^2$Department of Physics and Astronomy, University of Tennessee, Knoxville,
Tennessee, 37996 \\
$^3$Institute for Physics and Astronomy, University of Aarhus, Denmark \\
and Theoretical Center for Astrophysics, University of Aarhus, Denmark \\
$^4$ Cumberland University, Lebanon, Tennessee 37087 \\
$^5$Laboratory for Laser Energetics, University of Rochester, 
250 E. River Road, \\
Rochester, NY 14623
}
\date{\today}
\maketitle

\tightenlines

\begin{abstract}
We present Gamow-Teller strength distributions 
from shell model Monte Carlo 
studies of $fp$-shell nuclei that may play an important
role in the pre-collapse evolution of supernovae. 
We then use these strength distributions to calculate
the electron-capture cross sections and rates in the 
zero-momentum transfer limit. 
We also discuss the thermal behavior of the cross sections. 
We find large differences in these cross sections and rates
when compared to the
naive single-particle estimates. These differences
need to be taken into account for improved
modeling of the early stages of type II supernova evolution.

\end{abstract}

\pacs{PACS numbers: 26.50.+x, 23.40.-s, 21.60Cs, 21.60Ka}

\narrowtext

\section{Introduction}

The impact of nuclear structure on 
astrophysics has become increasingly 
important, particularly
in the fascinating and presently unsolved problem of type-II supernovae 
explosions. The possibility to detect neutrinos ejected before
the infalling matter reaches the 
neutrino trapping density\cite{sataria97} will allow us, for the
first time, to understand whether models of the precollapse
evolution are in reasonable agreement with observation. 
These studies will also shed light 
on the effective electron-to-proton ratio in the
early stages of the collapse. 

Key inputs for the precollapse scenario  are the electron capture
cross sections and rates for iron group nuclei 
\cite{Bethe90,Aufderheide}.
The core of a massive
star at the end of hydrostatic burning is stabilized by electron degeneracy
pressure as long as its mass does not exceed the appropriate Chandrasekhar
mass $M_{CH}$. If the core mass exceeds $M_{CH}$, electrons are captured by
nuclei \cite{Bethe90}. 
Thus, the depletion of the electron population due to
capture by nuclei is a crucial factor 
determining the initial collapse phase.

The reduction of the electroweak interaction
matrix element to the zero momentum transfer limit for the nuclear sector
leads directly to the Gamow-Teller operator
as a primary ingredient in electron-capture cross section calculations. 
The Gamow-Teller (GT) properties of nuclei in the iron region of the periodic
table are known to be crucial for supernova physics \cite{FFN}. 
For many of the nuclei that are present
in this early stage of the presupernova \cite{Aufderheide},
GT transitions contribute significantly to the electron
capture cross sections.
Due to insufficient experimental information, the GT$_+$
transitions, which change protons into neutrons, 
have so far been treated only qualitatively in presupernova
collapse simulations, assuming the GT$_+$ strength to reside in a single
resonance whose energy relative to the daughter ground state has been
parametrized phenomenologically \cite{FFN}; the total GT$_+$ strength has
been taken from the single particle model. Recent $(n,p)$ experiments
\cite{gtdata1,gtdata2,gtdata3,gtdata4,gtdata5}, 
however, show that the GT$_+$ strength is
fragmented over many states, while the total strength is significantly
quenched compared to the single particle model. A recent update of the
GT$_+$ rates for use in supernova simulations assumed a constant quenching
factor of 2 \cite{Aufderheide}, and included the known low-lying
transitions in the calculations of the rates. 

In the presupernova collapse, electron capture on $pf$-shell nuclei 
proceeds at temperatures between 300 keV and 800 keV \cite{Aufderheide}.
While presupernova collapse studies took into account the temperature
effects on the electron capture rate induced by changes 
in the electron distribution, potential effects on the Gamow-Teller
strength distribution have been neglected, and use was
made of the extreme
shifting assumption \cite{Aufderheide:1991} which states that the centroid
corresponding to each parent excited state is shifted upward in the
daughter nucleus by the excitation 
energy of the parent state. As experiments
measure only the ground state distribution, this hypothesis has
necessarily to be tested theoretically, and it has
been found approximately valid in a restricted shell model study
\cite{Aufderheide:1993a,Aufderheide:1993b}; 
however, the recent SMMC calculations of Ref. \cite{Radha:1997}
clearly indicate changes of the Gamow-Teller strength distribution with
temperature. Thus it appears desirable to take possible changes of the
Gamow-Teller strength distribution with temperature also into account
when calculating presupernova electron capture rates. We will
consistently do this in this paper. 

In contrast to previous studies, 
the calculations presented in this paper
solve the full shell model 
problem for the Gamow-Teller strength distributions in the $0\hbar\omega$
$fp$-shell space using a realistic residual interaction. We use the shell
model Monte Carlo (SMMC) approach to find the Gamow-Teller strength 
distributions. SMMC has the added advantage that it treats
nuclear temperature exactly. 
These distributions are then used to calculate the electron-capture 
cross section as a function of the incident energy of the electron. 
We also calculate the electron-capture rates assuming a relativistic
electron gas and a variety of electron chemical potentials. 
In Section II we discuss the methods used in these calculations, while
in Section III we present our results. A discussion and conclusion follow in
Section IV. 

\section{Theoretical background}

\subsection{Shell Model Monte Carlo techniques}

The SMMC method is based on a statistical formulation of the
nuclear many-body problem. In the finite-temperature
version of this approach, an observable is
calculated as the canonical expectation value
of a corresponding operator $\hat{\cal A}$ at a given nuclear temperature 
$T_N$,
and is given by~\cite{johnson:1992,lang:1993,ormand:1994,physrep:1997}
\begin{equation}
\langle \hat {\cal A}\rangle=
{{\rm Tr_A} [\hat {\cal A} {\hat U}]\over
{\rm Tr_A} [\hat U]},
\end{equation}
where $\hat U=\exp(-\beta \hat H)$ is the imaginary-time
many-body propagator,
${\rm Tr_A} \hat U$ is the canonical partition function
for $A$ nucleons, $\hat H$ is the shell model Hamiltonian,
and $\beta=1/T_N$ is the inverse temperature.

The SMMC canonical expectation values are based on
the discretization of the many-body propagator, $\exp(-\beta \hat{H})$,
into a finite number of imaginary time slices, $N_t$, each of duration
$\Delta\beta=\beta/N_t$. At each time slice the many-body propagator
is linearized via the Hubbard-Stratonovich
transformation~\cite{hubbard:1957,strato:1957};
observables are thus expressed as path integrals of one-body
propagators in fluctuating auxiliary fields. The integration
is carried out by a Metropolis random walk~\cite{Met:1953}.
To circumvent the sign problem encountered in the SMMC
calculations with realistic interactions,
we use the extrapolation procedure
outlined in Refs.~\cite{alhassid:1994,dean:1995}.

Yet another, but distinct, source of the sign problem
is an odd number of nucleons
in the canonical expectation values~\cite{physrep:1997}.
We overcome this problem by a number-projection technique,
first employed in
\cite{dean:1994} and subsequently used in
\cite{physrep:1997}, that allows us to extract information concerning
odd-$A$ nuclei from the neighboring even-even system.
The partition function for the odd-$A$ nucleus $A'$ is given by
\begin{equation}
Z_{A'}=\int {\cal D}\left[\sigma\right]G(\sigma){\rm Tr}_{A'}U_\sigma
=\int {\cal D}\left[\sigma\right]
G(\sigma){\rm Tr}_A 
U_\sigma \frac{{\rm Tr}_{A'}U_\sigma}{{\rm Tr}_{A}U_\sigma}\;,
\end{equation}
where we have introduced the trace over the even-even nucleus $A$. 
The integration is over all auxiliary field variables, $\sigma$, and
$G(\sigma)$ is a Gaussian factor that arises when the Hubbard-Stratonovitch
transformation is employed. We define
our Metropolis sampling 
weight as $W(\sigma)=G(\sigma){\rm Tr}_A U_\sigma$, and in the 
case of number projection, the Monte Carlo
sign becomes $\Phi(\sigma)={\rm Tr}_{A'}U_\sigma/{\rm Tr}_{A} U_\sigma$, which
is small, but fairly stable against fluctuations at least down to 
$T_N=$~0.8~MeV for realistic interactions. 
We project from the nearest even-even
system with $A'+1$ particles. 

The SMMC method is also used to calculate the response function
$R_{\cal A}(\tau)$ of an operator ${\cal A}$ at 
an imaginary-time $\tau$.
The response describes the 
dynamical behavior of the nucleus under the
influence of the operator, and 
contains information about the nuclear spectrum.
This is seen by using a spectral distribution of initial and final states
$\mid i\rangle$, $\mid f\rangle$ with energies $E_i$ and $E_f$ 
\begin{equation}
\label{response function}
R_{\cal A}(\tau)\equiv\langle \hat {\cal A}^\dagger(\tau) \hat {\cal
A}(0)\rangle =
{ {\rm Tr}_A[e^{-(\beta-
\tau) \hat H} \hat {\cal A}^\dagger e^{-\tau \hat H} \hat {\cal A}]
\over {\rm Tr}_A [e^{-\beta \hat H}]}
={\sum_{if} (2J_i+1) e^{-\beta E_i} e^{-\tau(E_f-E_i)} {\vert \langle f \vert
\hat{\cal A}
\vert i \rangle \vert}^2 \over \sum_i (2J_i+1) e^{-\beta E_i} }.
\end{equation}
Note that the total strength for the operator is given by $R(\tau=0)$.
The strength distribution
\begin{equation}
S_{\cal A}(E)={{\sum_{if} \delta(E-E_f+E_i) (2J_i+1)
e^{-\beta E_i} \vert \langle f \vert
\hat {\cal A} \vert i \rangle \vert^2} \over {\sum_i (2J_i+1) e^{-\beta E_i}}}
\label{eq4}
\end{equation}
is related to
$R_{\cal A}(\tau)$
by a Laplace Transform,
\begin{equation}
\label{laplace transform}
R_{\cal A}(\tau) = \int_{-\infty}^\infty S_{\cal A} (E) e^{-\tau E} dE.
\end{equation}
Note that  
here $E$ is the energy
transfer within the parent nucleus,
and that the strength distribution $S_{GT+}(E)$ has units of MeV$^{-1}$.

\subsection{Weak interaction cross sections in nuclei}

Electron capture by nuclei is mediated by 
the electro-weak charged current. At energies appropriate for 
the pre-collapse supernova, the W-boson propagator may be 
collapsed to point coupling and the weak vertex represented by the 
(V-A) coupling with the universal strength $g_{\rm wk}$.
The complexity of computing these channels 
for electron capture arises from the need to treat both the weak sector 
and the details of the nuclear structure 
effects with suitable finesse to obtain reliable
cross sections. 

Starting from the exact expression for the semi-leptonic process 
corresponding to electron capture, one can approximate the 
three-momentum transfer to the nucleus to be 
zero in the nuclear sector corresponding to computing the nuclear 
matrix elements between initial and final states. This 
allows us to calculate and predict 
the nuclear sector with a high level of accuracy. 
The lepton traces and 
kinematics are still included correctly as is energy 
conservation between the
initial and final states. The sacrifice of the magnitude of the 
momentum transfer to the nucleus in the nuclear sector
is reasonably 
well justified at the energy domain operative in the precollapse stellar 
core where energies are low and
the energy transfer from the incident charged
lepton to the nucleus is mainly used for internal nuclear excitation.

Electrons with energy $E_e$ 
interact with the nucleus through the weak interaction. 
For charge exchange interactions
\begin{equation}
e^- + A(N,Z) \rightarrow A(N+1,Z-1) + \nu_{e^-}\;.
\end{equation}
The parent nucleus absorbs a part of the incident energy and the
neutrino carries the remaining 
energy $\varepsilon_\nu$. The energy absorbed by the parent
is given by the difference between the initial and final nuclear states
$E=E_f-E_i$, and the states are labeled by 
$\{\mid f\rangle, \mid i\rangle\}$, respectively. Energy conservation 
requires that $E_e=E+\varepsilon_\nu$. 
The details of the derivation of the matrix element
are given in \cite{oconnell:1972}.
The angular cross section follows as
\begin{equation}
\frac{d\sigma}{d\Omega} =\frac{\varepsilon_\nu^2 V}{(2\pi)^2}\left[
\sum_{lepton spins} \frac{1}{2J_i+1}\sum_{M_f M_i} 
\mid \langle f \mid \hat{H}_W \mid i\rangle\mid^2\right]\;,
\end{equation}
where $\hat{H}_W$ is the weak interaction Hamiltonian, 
$J_i$ ($J_f$) is the initial (final) spin of the nuclear state, and
$M_i,M_f$ are projections of the angular momentum of the initial 
and final states of the nucleus. In the limit of small three-momentum
transfer, the partial cross section for electron capture between fixed
initial and final states  at a given
incident electron energy reduces to 
\begin{equation}
\sigma_{fi}(E_e)=
\frac{6(E_e-E)^2g_{\rm wk}^2}
{\pi(2J_i+1)} \mid\langle f \mid\mid \hat{\cal L}_1
\mid\mid i\rangle\mid^2
\label{sigma_discrete}
\end{equation}
where $g_{\rm wk}=1.1661\times10^{-5}$~GeV$^{-2}$ 
is the weak coupling constant, and
\begin{equation}
\hat{\cal L}_{1M} = 
\frac{i}{\sqrt{12\pi}}G_A \sum_{i=1}^A \tau_{+}(i)\sigma_{1M}(i)\;.
\end{equation}
Here $G_A$ is the axial vector form-factor which at zero momentum is
$G_A=1.25$. The Gamow-Teller operator is
${\bf GT}_{+}=\sum_i\tau_i^{+}{\bf \sigma}_i$, where $\tau_i^{+}$ 
is the isospin raising  operator, 
and ${\bf \sigma}_i$ is
the Pauli spin operator for nucleon $i$. The Clebsch-Gordon 
coefficients that arise when applying the Wigner-Eckart theorem give
selection rules for
this operator that 
require transitions from an initial state with quantum numbers
($J_i,T_i,\Pi_i$) to a final state ($J_f,T_f,\Pi_f$) such that
$\Delta J=J_f-J_i=0,1$, but no $J_f=J_i=0$, $\Delta T=T_f-T_i=0,1$, and
$\Pi_f=\Pi_i$. 

Our previous studies showed that the experimentally observed
quenching of the total GT strength is consistently reproduced by the
correlations within the full $pf$-shell if a renormalization of the spin
operator by the factor 0.8 is invoked \cite{karli:1995,caurier:1994}.
The same
renormalization factor had already been deduced from sd-shell
\cite{wildenthal:1988} and
$pf$-shell nuclei with $A \leq 49~$\cite{caurier2:1996,poves:1997}
and thus appears to be universal. In this work, we also renormalize the 
spin operator by 0.8, $\sigma_{\rm eff} = 0.8 \sigma_1$. 

In order to obtain the total cross section at
a given incident electron energy, one must sum over all initial
and final states in
Eq.~(\ref{sigma_discrete}). 
We are generally interested in situations when the initial nucleus
is at a finite temperature, so that the initial state sum needs
to be weighted by the appropriate statistical factors.
Thus, the total cross section is 
\begin{eqnarray}
\sigma(E_e) &=& \sum_{if}\frac{(2J_i+1)\exp(-\beta E_i)}{Z_A}
\sigma_{fi}(E_e)\; \\
&=& \frac{6 g^2_{\rm wk}}{\pi} \int d\xi (E_e - \xi)^2 \sum_{if}
\frac{\exp(-\beta E_i)}{Z_A} \frac{G_A^2}{12\pi} |{\langle f || \sum_{i=1}^A
\tau_+ (i)
\sigma_{\rm eff} (i) || i \rangle}|^2 \delta (\xi-E_f+E_i) \: .
\end{eqnarray}
Using the SMMC expression for the Gamow-Teller strength distributions
and shifting the energy scale in (4) by an appropriate Coulomb
correction and the proton-neutron mass splitting,
the total cross section can be recast into the form
\begin{equation}
\sigma(E_e)=
6g_{\rm wk}^2\int d\xi 
(E_e-\xi)^2\frac{G_A^2}{12\pi}
S_{GT_+}(\xi)F(Z,E_e)\;.
\end{equation}
Here we have additionally
accounted for the distortion of the
electron's wave function due to the Coulomb field of the nucleus by
introducing the Fermi function
$F(Z,E_e)=2\pi\eta/\left[\exp(2\pi\eta)-1\right]$ 
where $\eta=Z\alpha/\hat{\beta}$,
$\alpha$ is the fine-structure constant, 
$\hat{\beta}=\mid\vec{\beta}\mid$ is the magnitude of 
the velocity of the incoming
electron, and $Z$ is the charge of the parent nucleus. 

The presupernova electron capture rate $\lambda_{\rm ec}$ is then given
by folding the total cross section 
with the flux of a degenerate 
relativistic electron gas:
\begin{equation}
\lambda_{\rm{ec}}=\frac{\ln 2}{6163 \rm{sec}}\int_0^\infty
d\xi S_{\rm{GT}}(\xi,T_N) 
\frac{c^3}{(m_e c^2)^5}\int_{\cal L}^\infty dp p^2 (-\xi+E_e)^2
\frac{F(Z,E_e)}{1+\exp\left[\beta_e(E_e-\mu_e)\right]}\;,
\end{equation}
where $1/\beta_e$, $p$, and $\mu_e$ are the electron temperature,
momentum, and
chemical potential, and
${\cal L}=(Q_{if}^2 - m_e^2 c^4)^{1/2}$ for $Q_{if} \leq -m_e c^2$, and
0 otherwise, and $Q_{if}=-E=E_i-E_f$ is the  energy difference between
the nuclear states $i$ and $f$, respectively. 
We have indicated that the Gamow-Teller distribution is calculated at
the finite nuclear temperature 
$T_N$ which, in principle, is the same as the one of the electron
gas $T_e=1/\beta_e$. 
However, we will study below the uncertainties introduced in the electron
capture rate due to the temperature dependence of the 
Gamow-Teller distribution by varying $T_N$ for fixed $T_e$.

\section{Results}

\subsection{Validation}

The SMMC calculations of Gamow-Teller strength distributions were detailed in
Ref.~\cite{Radha:1997}. In the present work, we have calculated several
more nuclei in addition to those found in Ref. \cite{Radha:1997}. Each
even-even nucleus is calculated at a temperature of $T_N=0.5$~MeV, while
the $g$-extrapolation required to circumvent the SMMC sign problem
allows only to cool
odd-$A$-nuclei down to a temperature of $T_N=0.8$~MeV. 
We use a $\Delta\beta=0.03125$~MeV$^{-1}$, 
and approximately four-thousand samples per extrapolation Hamiltonian. 
We use the KB3 residual interaction\cite{kb3:1981} which is well
suited for full $0\hbar\omega$ studies throughout the lower $pf$-shell
region\cite{caurier:1994}. Using the numerical techniques described
in Ref.~\cite{Radha:1997,physrep:1997}, we construct a strength distribution
from the response function of the Gamow-Teller operator.
Recall from Ref.~\cite{Radha:1997} that good agreement was found between 
experiment and theory for those nuclei that had been experimentally 
investigated. 

In order to demonstrate and validate that 
our SMMC odd-$A$ projection procedure 
gives reasonable results for the strength
distributions at finite temperatures,
we compare our SMMC results in $^{45}$Sc to standard shell model 
diagonalization calculations. $^{45}$Sc does not actually play a role
during the presupernova collapse. Its choice is, however, dictated as it
is the heaviest odd-$A$  nucleus for which state-by-state calculations of the
Gamow-Teller strength distribution at finite temperature in the complete
$pf$-shell are feasible. 
Using Eq. 3 above, we calculated $S_{GT_+}(E,T_N)$ within the
diagonalization approach for
all initial states in $^{45}$Sc 
up to an excitation energy of 4.68~MeV. At a nuclear temperature of 
0.8~MeV, the last excited state considered contributes less than
0.09\% to the partition function. Figure 1 compares the thermal 
distribution (dotted histogram) obtained by diagonalization with the
SMMC result (solid line) calculated at the same temperature $T_N=0.8$
MeV. The distributions are plotted as function of the nuclear energy transfer
$E=E_f-E_i$ required in (12). (The Coulomb energy of the final state has been
corrected using the shift as defined in \cite{karli:1995}.)   
One observes that the SMMC calculation reproduces both the position
and the width of the thermal distribution very well.
The diagonalization calculation gives a total
strength of 0.60, while the 
SMMC result is 0.52$\pm$0.08.

To study the temperature dependence of the Gamow-Teller strength
distribution we have performed the diagonalization calculations at different
temperatures $T_N=0.$, 0.3 MeV and 0.8 MeV The results are also given in
Fig. 1. At first we note that the $T=0$ distribution can be compared to the
experimental data of Ref. \cite{sc45_expt}. The total calculated
strength (0.52$\pm$0.08) 
agrees well with the measurement (0.48 up to excitation
energies of 9 MeV; the observed strength at higher excitation energies
is likely due to contributions outside the present model space
\cite{caurier2:1996}). However, the Gamow-Teller peak of the 
experimental distribution is
about 1 MeV higher in energy than the calculation, which makes the
agreement between data and shell model calculation worse than in the
other $pf$-shell nuclei \cite{caurier:1994,Radha:1997}.
With increasing temperature the Gamow-Teller distributions are broadened
and its center is shifted slightly to lower transfer energies. Both
effects have already been noticed in the SMMC studies of
\cite{Radha:1997}. The broadening is mainly due to the fact that the
number of states contributing to the thermal ensemble increases. The
energy shift is caused by the decrease of pairing energy in the higher
excited states. We mention that the large peak seen in the
distributions at $T_N=0$ and 0.3 MeV 
corresponds principally to $^{45}$Sc$(J=7/2)\rightarrow ^{45}$Ca$(J=5/2)$
transitions. In the distribution at $T_N=0.8 $ MeV this peak appears 
lower in energy at $E\approx 5$ MeV and its strength is noticeably
reduced.

We also want to understand how 
the electron capture cross section (11), as calculated with the
SMMC Gamow-Teller strength distribution at $T_N=0.8$ MeV, compares to the
one obtained from the respective diagonalization results. The middle plot
in Fig. 1 shows 
that the SMMC result reproduces the thermal diagonalization data down
to $E_e=10$~MeV very closely, and is within 30\% of the exact result
down to $E_e=5$~MeV. At very low electron energies the cross section
becomes very sensitive to the discrete level structure of the 
Gamow-Teller transitions. These weak transitions are not well reproduced
by the SMMC calculation leading to a noticeable underestimate of the
cross section within the SMMC approach. 
However, the cross section is already very low at these electron
energies and, as we will see next, this shortcoming of the SMMC
calculation is not too important for the electron capture rate.
Note that the standard shell model calculations yield almost
oscillatory behavior at low $E_e$ (dashed line), which reflects its
discrete level structure at low excitation energies.
For completeness, we have also plotted the electron capture cross
sections calculated for the Gamow-Teller distributions at $T_N=0$ and
$T_N=0.3$ MeV. We observe that differences in the distribution only
affect the cross section at rather low electron energies.

The quantity of interest for presupernova studies is the electron
capture rate (12).  In the lower panel of Fig. 1 we compare the rates, 
calculated from the SMMC and from the diagonalization Gamow-Teller
strength distributions at $T_N=0.8$ MeV, as a function of electron chemical
potential.  The agreement is very
satisfying for the entire range $\mu_e \le 10$ MeV indicating that the
differences between the SMMC and diagonalization electron capture cross
sections at low electron energies are washed out when folded with the
electron flux.

As mentioned above, SMMC calculations with realistic interactions 
cannot be performed for odd-$A$ nuclei at temperatures below $T_N=0.8$
MeV. As the relevant temperature during the early stage of the 
presupernova collapse is lower, we have investigated the inaccuracy 
introduced in the electron capture rates if the Gamow-Teller distribution
at these lower temperatures is approximated by the one at $T_N=0.8$ MeV.
Choosing the extreme case, $T_N=0.3$, we have performed one consistent
calculation of the electron capture rate using the appropriate electron
flux and Gamow-Teller strength distribution (determined within the
diagonalization approach). This result is compared with a calculation in
which the Gamow-Teller distribution is approximated by the SMMC
distribution calculated at $T_N=0.8$ MeV. We observe a maximal difference
of a factor 4 for $\mu_e < 10$ MeV which mainly reflects deviations in
the Gamow-Teller distribution at low energies. Thus, it can be reduced
if the weak discrete transitions which are experimentally known are added by
hand to the SMMC results, following the procedure of \cite{FFN}.
Obviously the present SMMC approach becomes increasingly more accurate
for later stages of the collapse where the temperature is higher and for
even-even parent nuclei where SMMC calculations at the relevant
temperature are possible.

Finally, we want to study how much the statistical Monte Carlo
uncertainties and the errors related to the inversion
technique used to determine the Gamow-Teller distribution from the
response function (5) affect the electron capture rates. The major
uncertainty introduced in our Gamow-Teller strength distribution is in
the position of the centroid which has an error of about $\pm 0.5$ MeV.
In Fig. 2 we show how much such a change would affect
$\lambda_{\rm ec}$.
Decreasing (increasing) the position of the centroid by 0.5~MeV increases 
(decreases) the rate by an approximate factor of 1.4 (1.3).

\subsection{Electron capture results}

Ref. \cite{Radha:1997} presented SMMC Gamow-Teller distributions for those
$pf$-shell nuclei in the iron mass region for which experimental data
are available ($^{54,56}$Fe, $^{58,60,62,64}$Ni, $^{51}$V,
$^{55}$Mn and $^{59}$Co). The agreement between theory and data was
very satisfying in all cases, and it was concluded that the SMMC
approach has the predictive power to estimate the Gamow-Teller strength
distribution for nuclei for which no data exist. We have extended the
study of Ref. \cite{Radha:1997} by performing SMMC
calculations for several of those $pf$-shell nuclei which are of
interest for the presupernova collapse: $^{55,57}$Co, $^{56}$Ni,
$^{50,52}$Cr, $^{55,58}$Fe, $^{56}$Ni, and $^{50}$Ti. 
The Gamow-Teller distributions for these nuclei 
are plotted in Fig.~3 as
a function of energy transfer $E=E_f-E_i$. As for $^{45}$Sc,
the Coulomb correction has been performed following the prescription
given in Ref.~\cite{karli:1995}.
Each Gamow-Teller distribution has been interpolated onto a fine energy
grid with a spline function. The spline function was
then used to calculate the cross section using 
the formulae described in Section II. 

Before presenting results for the electron capture cross sections and
rates, we note that our approach, as presented in Section
II, implicitly assumes that the calculation reproduces the mass
splitting between parent and daughter nucleus correctly. As has been
shown in \cite{caurier:1994,karli:1995} the KB3 interaction indeed
describes masses of $pf$-shell nuclei reasonably well (typically within
500 keV of the data; as the deviations have a systematical trend the
inaccuracy on the relevant mass splitting between parent and daughter
should be even smaller). Many of the SMMC masses presented in
\cite{karli:1995} have very recently also been calculated within the
conventional diagonalization approach, and the SMMC results (after
subtraction of the appropriate internal excitation energy) have been
confirmed in all cases \cite{Gabriel:1998}. These studies, however,
clearly showed \cite{karli:1995,Gabriel:1998} that the KB3
interaction overbinds the $f_{7/2}$ subshell closure. This shortcoming
can be circumvented \cite{Otsuka:1998} 
by using the FDP6 interaction \cite{richter:1991}. 
However, this force
does not reproduce the Gamow-Teller strength correctly \cite{Dean:1994}. 
Weighing the advantages and disadvantages of the two interactions, we have
chosen the KB3 interaction with the price that we had to shift the
energy scale of the Gamow-Teller distributions in order to correct for
the $f_{7/2}$ subshell overbinding. 
As can be seen in Ref. \cite{Radha:1997}, after correction the SMMC
calculation reproduces the observed Gamow-Teller distributions in
$^{54}$Fe and the various Ni isotopes very well.

In Table 1 we compare the SMMC results for the Gamow-Teller peak in the
daughter nucleus with the parametrization given by Fuller, Fowler, and
Newman \cite{FFN} and subsequently used in the  update of the
presupernova electron capture rates by Aufderheide {\it et al.}
\cite{Aufderheide}.
We observe that Ref. \cite{FFN} generally places the Gamow-Teller
strength for even-even parent nuclei at too high an excitation energy in
the daughter, while it is the opposite for odd-$A$ parent
nuclei. The same trend as in the SMMC distributions
is found in the data and has already been
pointed out in Ref. \cite{Koonin:1994}.
From the SMMC calculations and the data, we can conclude that
for an even-even parent, the
Gamow-Teller strength, at low temperatures, is at low daughter 
excitation energies ($\sim 2$ MeV),
while it is at significantly higher excitation energies ($\sim 5$ MeV)
for an odd-$A$ nucleus. This different behavior is related to the pairing
energy difference between the even-even parent and odd-odd daughter.
As the Gamow-Teller distributions usually have widths of order 1-2 MeV,
our SMMC calculations for odd-$A$ nuclei 
might miss weak Gamow-Teller transitions at low excitation energies 
which could potentially increase the electron capture rates. It appears
therefore reasonable to follow the prescription of \cite{FFN} and add
experimentally known transitions at low energies to the rates obtained
from the SMMC
Gamow-Teller distributions for odd-$A$ nuclei. Such a procedure seems to
be unnecessary for even-even parent nuclei.

The electron capture cross sections calculated from the 
Gamow-Teller matrix elements are presented in Fig.~3 as a function of
the incident electron energy $E_e$.
The general behavior of the cross section can be easily understood. To
trigger the electron capture process requires a minimum electron energy
given by the mass splitting between parent and daughter, $Q_{if}$.  (This
threshold is lowered by the internal excitation energy at finite
temperature.) In
even-even parent nuclei the Gamow-Teller strength, at low temperatures, 
is centered at daughter
excitation energies of order 2 MeV. Thus the electron capture cross
sections for these parent nuclei increase drastically within the first
couple of MeV of electron energies above threshold, 
reflecting the Gamow-Teller distribution. With
increasing electron energies it continues to raise modestly caused by
the $(E_e-\xi)^2$ factor in the cross section integral (12). 
As $Q_{if}$
increases with neutron excess, caused by the increase of the symmetry
energy,
electron capture cross sections, at fixed electron energies, decrease
with neutron excess. 
In odd-$A$ nuclei the Gamow-Teller distribution, at low
temperatures, peaks at noticeably higher daughter
excitation energies. Thus the electron capture cross sections are
shifted to higher electron energies for odd-$A$ nuclei in comparison to
even-even parent nuclei by about 3 MeV, reflecting the difference in
the Gamow-Teller peak positions.

Shown in Fig.~5 are the electron capture rates 
as a function of the electron chemical potential.   
We have used $T_e=T_N=0.5$~MeV for the electron capture rates on even
parent nuclei, while for odd-$A$ parent nuclei we used $T_N=0.8$ MeV for
the reasons discussed above.
The electron chemical potential depends on the electron density, the
electron fraction in the medium, and the temperature of the electrons.
In the precollapse phase of supernova, a reasonable approximation is given
by \cite{BR78}
\begin{equation}
\mu_e = 1.11 (\rho_7 Y_e)^{1/3}\left[1+\left(\frac{\pi}{1.11}\right)^2
\frac{T_e^2}{\left(\rho_7 Y_e\right)^{2/3}}\right]^{-1/3}\;,
\end{equation}
where $T_e$ is in units of MeV, and the electron density $\rho_7=\rho/10^7$. 
Thus, for the nuclei studied here $\mu_e \approx 1.5-2$ MeV under
typical presupernova conditions ($T_N\approx 0.4 $ MeV, $\rho_7 \approx
6$).

Do the present electron capture rates indicate potential implications for
the pre-collapse  evolution of a type II supernova? To make a 
judgement on this
important question, we compare in Table II the SMMC rates for selected
nuclei with those currently used in collapse calculations
\cite{Aufderheide}. For the comparison we choose the same physical
conditions as assumed in Tables 4--6 in \cite{Aufderheide}. Furthermore
the table lists the partial electron capture rate which has been
attributed to Gamow-Teller transitions in Ref. \cite{Aufderheide}.
At first we note that for even parent nuclei, the present rate
approximately agrees with the currently recommended {\it total} rate. A
closer inspection, however, shows significant differences between the
present rate and the one attributed to the Gamow-Teller transition in
\cite{Aufderheide}.  As discussed above, the origin of this discrepancy
is due to the fact that Ref. \cite{FFN} places the Gamow-Teller
resonance for even-even nuclei systematically at too high an excitation
energy. Of course, this shortcoming has been corrected for 
in Refs. \cite{FFN,Aufderheide} by adding an experimentally known 
low-lying strength on top
of the one attributed to Gamow-Teller transitions. 
However, the overall good agreement between the SMMC results for
even-even nuclei and the recommended rates indicates that our SMMC
approach also accounts correctly for this low-lying strength. This has already
been deduced from the good agreement between SMMC Gamow-Teller
distributions and data including the low-energy regime. We conclude
therefore that, for even-even nuclei, the SMMC approach is able to
predict the {\it total} electron capture rate rather reliably, even if
no experimental data are available. We note that our SMMC rate is
somewhat larger than the recommended rate for $^{56}$Fe and $^{60}$Ni.
In both cases the experimental Gamow-Teller distribution is known and
agrees well with the SMMC results \cite{Radha:1997}. While the proposed
increase of the rate for $^{60}$Ni is not expected to have noticeable
influence on the pre-collapse evolution, the increased rate for $^{56}$Fe
makes this nucleus an important contributor in the change of $Y_e$
during the collapse (see Table 15 of \cite{Aufderheide}).

Turning now to electron capture on odd-$A$ nuclei. Here we observe that
the SMMC electron capture rate, derived from the Gamow-Teller
distributions, is significantly smaller than the recommended total rate.
As already stressed above, this is simply due to the fact that for 
odd-$A$ nuclei the Gamow-Teller transition peaks at rather high
excitation energies in the daughter nucleus. The electron capture rate
on odd-$A$ nuclei is therefore carried by weak transitions at low
excitation energies. Comparing our rates to those attributed to
Gamow-Teller transitions in Refs. \cite{FFN,Aufderheide} reveals that 
the latter have been, in general, significantly overestimated which is
simply caused by the fact that the position of the Gamow-Teller
resonance is usually put at too low excitation energies in the daughter
(see Table I).
For example, Ref. \cite{Aufderheide} attributes $87\%$ of the total
capture rate on $^{55}$Co to Gamow-Teller transitions (upper part of
Table II), while our calculation predicts the contribution of the
Gamow-Teller strength distribution to be negligible. Thus, we recommend
that the capture rate on $^{55}$Co is significantly smaller (only $13\%$
of the rate given in Table 4 of Ref. \cite{Aufderheide}). 
Our SMMC calculation implies that the Gamow-Teller transitions should not
contribute noticeably to the electron capture rates on odd-$A$ nuclei at
the low temperatures studied in Tables 14--16 in \cite{Aufderheide}.
Thus, the
rates for odd-$A$ nuclei given in these tables should generally 
be replaced by the
non-Gamow-Teller fraction.

\section{Discussion}

In this paper we presented a detailed description of the 
electron-capture cross section and rates 
calculated from the Gamow-Teller distributions
obtained using the SMMC method. We demonstrated the validity of
our odd-$A$ sampling technique in $^{45}$Sc, and we found very good
agreement between the SMMC and the traditional approach for solutions of
the shell model. Furthermore, the SMMC approach reproduces the measured
Gamow-Teller strength distributions very well. In accord with the data,
we find an odd-even dependence of the Gamow-Teller peak position in the
daughter nucleus: while it is generally at low daughter excitation energy for
even-even parent nuclei ($\sim 2$ MeV), it is at noticeable higher
energies for odd-$A$ parents ($\sim 5$ MeV). These systematics are not
reproduced by the parametrization of the Gamow-Teller resonance as
adopted for the derivation of the currently recommended rates
\cite{FFN,Aufderheide}. This parametrization places the Gamow-Teller
resonance usually at too high excitation energies for even-even parent
while the position is too low for odd-$A$ parents. However, this
shortcoming has been mainly overcome in the recommended rates by adding
experimentally known strength at low excitation energies to the
Gamow-Teller strength.

The presupernova collapse occurs at finite temperature and our SMMC
approach, for the first time, allows to take thermal effects
consistently into account. With increasing temperature, the
Gamow-Teller distribution is broadened and shifted to lower transfer
energies. 
However, at the rather low temperatures at which electron capture occurs
on nuclei in the iron mass region, the temperature dependence of the
Gamow-Teller strength distribution is rather unimportant. For even-even
nuclei the distribution does not change too much at $T_N < 0.6$ MeV 
due to the large pairing gap. For odd-$A$ nuclei, the Gamow-Teller
strength resides at such high excitation energies that the modest
temperature dependence of the strength does not affect the total rate.
Thus it also does not matter that numerical problems do not allow us
to cool odd-$A$ nuclei below $T_N=0.8$ MeV. 
However, we like to stress that the temperature dependence of the
Gamow-Teller strength will become important at later stages of the
presupernova collapse when temperatures of order $T_N=1$ MeV or higher
are reached, as then the strength for both even-even and odd-$A$ nuclei
is noticeably shifted downwards in transfer energy. At these temperatures the
$Y_e$ value, however, has decreased enough so that electrons are
captured on nuclei with  $Z<40$ and $N>40$ for which the
Gamow-Teller transitions are naively expected to be Pauli-blocked.
However, this Pauli-blocking can be overcome by thermal excitation
\cite{Wambach:1984} and by pairing \cite{Langanke:1998}, which, at low
temperatures, is the more important effect.

As the Gamow-Teller strength resides at rather high excitation energies
for odd-$A$ parents, the Gamow-Teller contribution is generally
negligible in the total electron capture rate. This finding is at
variance with the recent compilations \cite{FFN,Aufderheide}
which propose noticeable
Gamow-Teller fractions for the capture 
rate on odd-$A$ nuclei like $^{55,59}$Co.
We recommend to use only the non-Gamow-Teller fraction of the compiled
rates for odd-$A$ nuclei. For even-even parents, the Gamow-Teller
distribution is located at very low excitation energies and, within our
approach, should account for the total electron capture rate. In fact,
our rate agrees reasonably well with the compiled total rates. However,
for nuclei like $^{56}$Fe and $^{60}$Ni, the Gamow-Teller strength
resides, in agreement between data and SMMC calculation
\cite{Radha:1997}, at such low energies that the present rate is
significantly larger than the one given in the compilations.

In summary, the present SMMC approach allows for the first time a
microscopic determination of the Gamow-Teller contributions to the
presupernova electron capture rates. The present calculation will be
extended to other nuclei of potential importance during the collapse
phase of type II supernova. The SMMC approach can also be extended to
heavier, more neutron-rich nuclei which will carry the electron capture
process at later stages. For these nuclei we expect the ability of our
method to consistently account for finite temperature effects to be
quite important.

\acknowledgements

This work was supported in part through grant DE-FG02-96ER40963
from the U.S. Department of Energy.
Oak Ridge National Laboratory (ORNL) is managed by Lockheed Martin Energy
Research Corp. for the U.S. Department of Energy under contract number
DE-AC05-96OR22464. 
KL has been partly supported by the Danish Research Council.
Grants of computational resources were provided by the Center
for Advanced Computational Research at Caltech, the
Center of Computational Science at ORNL, and NERSC.

\newpage
\begin{table}
\caption
{Comparisons of the positions of the Gamow-Teller peaks as calculated
within the SMMC approach with the parametrization of Ref. \protect\cite{FFN}.
The proton and neutron numbers refer to the parent nucleus, while $E_{GT}$
is the position of the Gamow-Teller peak in the daughter nucleus. (To
obtain this number we have shifted the transferred energy $E$ by the
experimental mass splitting between parent and daughter nucleus.)
$E_{\rm FFN}$ denotes the position of the Gamow-Teller resonance due to
the parametrization given in Ref. \protect\cite{FFN} and subsequently used in
presupernova collapse studies. The SMMC calculations have been performed
at $T_N=0.5 $ MeV for even nuclei and at $T_N=0.8$ MeV for odd-$A$
nuclei. For $^{50}$Cr the single-particle model allows transitions into
the $f_{7/2}$ and $f_{5/2}$ orbitals. 
}
\begin{tabular}{|c|c|c|c|}
N  &  Z & $E_{GT}$ & $E_{\rm FFN}$ \\
\hline
28 & 28 &   2.6  &  3.78 \\
30 & 28 &   2.0  &  3.76 \\
32 & 28 &   1.0  &  2.0  \\ 
28 & 27 &   6.9  &  3.78 \\
30 & 27 &   4.8  &  3.77 \\
32 & 27 &   4.2  &  2.0  \\
28 & 26 &   2.8  &  3.80 \\
29 & 26 &   6.1  &  5.4  \\
30 & 26 &   1.5  &  3.78 \\
32 & 26 &   -0.5 &  2.0  \\
30 & 25 &   4.7  &  3.79 \\
26 & 24 &   2.5  &  2.0,8.7  \\
28 & 24 &   1.3  &  3.82 \\
28 & 23 &   5.5  &  3.83 \\
\end{tabular}
\end{table}

\newpage
\begin{table}
\caption
{Comparisons of the present SMMC electron capture rates  with the total 
($\lambda_{\rm ec}$) and
partial Gamow-Teller ($\lambda^{\rm GT}_{\rm ec}$) rates as given in 
Ref.~\protect\cite{Aufderheide}. Physical conditions at
which the comparisons were made are given in the last column.
}
\begin{tabular}{|c|c|c|c|c|}
nucleus 
& $\lambda_{\rm ec}$ (sec$^{-1}$) 
& $\lambda_{\rm ec}$ (sec$^{-1}$) 
& $\lambda_{\rm ec}^{GT}$ (sec$^{-1}$) 
& conditions \\  
&   (SMMC)  &  
(Ref.~\protect\cite{Aufderheide}) & 
(Ref.~\protect\cite{Aufderheide}) &  \\
\hline 
$^{55}$Co &  2.25E-04  &   1.41E-01 & 1.23E-01 
& $\rho_7=5.86$, $T_9=3.40$, $Y_e=0.47$ \\
$^{57}$Co &  1.97E-06  &   3.50E-03 & 1.31E-04 
& $\rho_7=5.86$, $T_9=3.40$, $Y_e=0.47$ \\
$^{54}$Fe &  4.64E-05  &   3.11E-04 & 9.54E-07 
& $\rho_7=5.86$, $T_9=3.40$, $Y_e=0.47$ \\
$^{55}$Fe &  7.22E-09  &   1.61E-03 & 1.16E-07 
& $\rho_7=5.86$, $T_9=3.40$, $Y_e=0.47$ \\
$^{56}$Ni &  1.96E-02  &   1.60E-02 & 6.34E-03 
& $\rho_7=5.86$, $T_9=3.40$, $Y_e=0.47$ \\
$^{58}$Ni &  6.02E-04  &   6.36E-04 & 4.04E-06 
& $\rho_7=5.86$, $T_9=3.40$, $Y_e=0.47$ \\
$^{60}$Ni &  1.34E-05  &   1.49E-06 & 4.86E-07 
& $\rho_7=5.86$, $T_9=3.40$, $Y_e=0.47$ \\
\hline
$^{59}$Co &  2.05E-07  &   2.09E-04 & 6.37E-05 
& $\rho_7=10.7$, $T_9=3.65$, $Y_e=0.455$ \\
$^{57}$Co &  1.21E-05  &   7.65E-03 & 3.69E-04 
& $\rho_7=10..7$, $T_9=3.65$, $Y_e=0.455$ \\
$^{55}$Fe &  6.48E-08  &   3.80E-03 & 5.51E-07 
& $\rho_7=10.7$, $T_9=3.65$, $Y_e=0.455$ \\
$^{56}$Fe &  5.86E-06  &   4.68E-07 & 6.60E-10 
& $\rho_7=10.7$, $T_9=3.65$, $Y_e=0.455$ \\
$^{54}$Fe &  2.26E-04  &   9.50E-04 & 3.85E-06 
& $\rho_7=10.7$, $T_9=3.65$, $Y_e=0.455$ \\
$^{51}$V  &  6.17E-07  &   1.24E-05 & 9.46E-09 
& $\rho_7=10.7$, $T_9=3.65$, $Y_e=0.455$ \\
$^{52}$Cr &  5.22E-07  &   2.01E-07 & 1.59E-10 
& $\rho_7=10.7$, $T_9=3.65$, $Y_e=0.455$ \\
$^{60}$Ni &  7.48E-05  &   7.64E-06 & 2.12E-06 
& $\rho_7=10.7$, $T_9=3.65$, $Y_e=0.455$ \\
\end{tabular}
\end{table}

\begin{figure}
\caption{Top panel: The Gamow-Teller transitions
as a function of the energy transfer $E=E_f-E_i$
using SMMC 
for $^{45}$Sc, and comparing to 
diagonalization calculations (DD) at $T_N=0.33$ and $0.8$~MeV. 
Middle panel: The cross section calculated from both SMMC
and diagonalization. Bottom panel: 
Calculated rates from SMMC ($T_N=0.8$~MeV),
assuming $T_e=0.3$ and $0.8$~MeV, 
compared to the exact results ($T_e=T_N$).}
\label{fig1}
\end{figure}

\begin{figure}
\caption{
The error band in the rates calculations for $^{45}$Sc, where the
position of the calculated Gamow-Teller centroid has been shifted
by 0.5~MeV.} 
\label{fig2}
\end{figure}

\begin{figure}
\caption{Shown are the $GT_+$ distributions for various nuclei studied
here as a function of the energy transfer from the electron,
$E=E_f-E_i$.
These curves were obtained after MaxEnt reconstruction of the 
Gamow-Teller response functions that were 
calculated in the SMMC framework.}
\label{fig3}
\end{figure}

\begin{figure}
\caption{Shown are the electron capture
cross sections as a function of the incident electron
energy $E_e$ for all nuclei in this study.}
\label{fig4}
\end{figure}

\begin{figure}
\caption{Shown are the electron capture rates
as a function of the electron chemical potential
energy $\mu_e$ for all nuclei in this study. All electron
temperatures were fixed at $T_e=0.5$~MeV.}
\label{fig5}
\end{figure}


\begin{references}

\bibitem{sataria97}
F.K. Sataria and A. Ray, Phys. Rev. Lett. {\bf 79}, 1599 (1997)

\bibitem{Bethe90} 
H.A. Bethe, Rev. Mod. Phys. {\bf 62} 801 (1990)

\bibitem{Aufderheide} 
M. B. Aufderheide, I. Fushiki, S. E. Woosley, and D. H.
Hartmann, Astrophys. J. Suppl. 91, 389 (1994) 

\bibitem{FFN} 
G.M. Fuller, W.A. Fowler and M.J. Newman, ApJS {\bf 42}, 447 (1980);
{\bf 48}, 279 (1982); ApJ {\bf 252} (1982) 715; {\bf 293}, 1 (1985)

\bibitem{gtdata1} 
A.L. Williams {\it et al.}, Phys. Rev. {\bf C51}, 1144 (1995) 

\bibitem{gtdata2} 
W.P. Alford {\it et al.}, Nucl. Phys. {\bf A514}, 49 (1990) 

\bibitem{gtdata3} 
M.C. Vetterli {\it et al.}, Phys. Rev. {\bf C40}, 559 (1989)

\bibitem{gtdata4} 
S. El-Kateb {\it et al.}, Phys. Rev. {\bf C49}, 3129 (1994)

\bibitem{gtdata5} 
T. R\"onnquist {\it et al.}, Nucl. Phys. {\bf A563}, 225 (1993)

\bibitem{Aufderheide:1991}
M.B. Aufderheide, Nucl. Phys. {\bf A526}, 161 (1991)

\bibitem{Aufderheide:1993a}
M.B. Aufderheide, S.D. Bloom, D.A. Ressler and G.J. Mathews,
Phys. Rev. {\bf C47}, 2961 (1993)

\bibitem{Aufderheide:1993b}
M.B. Aufderheide, S.D. Bloom, D.A. Ressler and G.J. Mathews,
Phys. Rev. {\bf C48}, 1677 (1993)
 
\bibitem{Radha:1997}
P.B. Radha, D.J. Dean, S.E. Koonin, K. Langanke, and P. Vogel,
Phys. Rev. {\bf C56}, 3079 (1997)

\bibitem{johnson:1992}
C.~W. Johnson, S.~E. Koonin, G.~H. Lang, and W.~E. Ormand, {Phys. Rev.. Lett.}
{\bf 69},  3157  (1992)

\bibitem{lang:1993}
G.~H. Lang, C.~W. Johnson, S.~E. Koonin, and W.~E. Ormand, {Phys. Rev. C} {\bf
  48},  1518  (1993)

\bibitem{ormand:1994}
W.~E. Ormand, D.~J. Dean, C.~W. Johnson, G.~H. Lang, and S.~E. Koonin, {Phys.
  Rev. C} {\bf C49},  1422  (1994)

\bibitem{physrep:1997}
S.~E. Koonin, D.~J. Dean, and K. Langanke, {Phys. Rep.} {\bf 278}, 1 (1997)

\bibitem{hubbard:1957}
J. Hubbard, {Phys. Rev. Lett.} {\bf 3},  77  (1959)

\bibitem{strato:1957}
R. Stratonovich, {Dokl. Akad. Nauk. SSSR} {\bf 115},  1097  (1957)

\bibitem{Met:1953}
N. Metropolis, A. Rosenbluth, M. Rosenbluth, A. Teller, and E. Teller,
J. Chem. Phys. {\bf 21}, 1087 (1953)

\bibitem{alhassid:1994}
Y. Alhassid, D.~J. Dean, S.~E. Koonin, G. Lang, and
W.~E. Ormand, {Phys. Rev.  Lett.} {\bf 72},  613  (1994)

\bibitem{dean:1995}
D.~J. Dean, S.~E. Koonin, K. Langanke, P.~B. Radha, and Y. Alhassid,
{Phys.  Rev. Lett} {\bf 74},  2909  (1995)

\bibitem{dean:1994}
D.J. Dean, B.P. Radha, K. Langanke, S.E. Koonin, Y. Alhassid, and
W.E. Ormand, Phys. Rev. Lett, {\bf 72}, 4066 (1994)

\bibitem{oconnell:1972}
J.S. O'Connell, T.W. Donnelly, and J.D. Walecka,
Phys. Rev. {\bf C6}, 719 (1972)

\bibitem{karli:1995}
K. Langanke, D.~J. Dean, P.~B. Radha, Y. Alhassid, and S.~E. Koonin,
{Phys. Rev. C} {\bf 52},  718  (1995)

\bibitem{caurier:1994}
E. Caurier, A. Zuker, A. Poves, and G. Martinez-Pinedo, {Phys. Rev. C} {\bf
 50},  225  (1994)

\bibitem{wildenthal:1988}
B. Brown and B. Wildenthal, {Ann. Rev. Nucl. Part. Sci.} {\bf 38},  29
(1988)

\bibitem{caurier2:1996}
G. Martinez-Pinedo, A. Poves, E. Caurier, and A.~P. Zuker, {Phys. Rev. C}
{\bf 53},  R2602  (1996)

\bibitem{poves:1997}
G. Martinez-Pinedo, A.P. Zuker, A. Poves, and
E. Caurier, Phys. Rev. {\bf C55}, 187 (1997)

\bibitem{kb3:1981}
A. Poves and A. Zuker, {Phys. Rep.} {\bf 70},  235  (1981)


\bibitem{kbrown:1968}
T.~T.~S. Kuo and G.~E. Brown, {Nucl. Phys. A} {\bf 114},  241  (1968)

\bibitem{sc45_expt}
W.P. Alford {\it et al}, Nucl. Phys. {\bf A531}, 97 (1991)

\bibitem{Gabriel:1998}
G. Martinez-Pinedo, private communication

\bibitem{Otsuka:1998}
T. Otsuka, M. Honma and T. Mizusaki,
Phys. Rev. Lett., in print

\bibitem{richter:1991}
W. A. Richter, M. G. Vandermerwe, R. E. Julies, and B. A.
Brown, Nucl. Phys. {\bf A523} (1991) 325

\bibitem{Dean:1994}
D.J. Dean, B.P. Radha, K. Langanke, S.E. Koonin, Y. Alhassid, and
W.E. Ormand, Phys. Rev. Lett, {\bf 72}, 4066 (1994)


\bibitem{Koonin:1994}
S.E. Koonin and K. Langanke, Phys. Lett {\bf B326}, 5  (1994)

\bibitem{BR78}
S.A. Bludman and K.A. van Riper, Astrophys. J. {\bf 224}, 631 (1978)

\bibitem{Wambach:1984} 
J. Cooperstein and J. Wambach, Nucl. Phys. {\bf A420}, 591 (1984)

\bibitem{Langanke:1998}
K. Langanke, E. Kolbe and D.J. Dean, in preparation

\end{references}
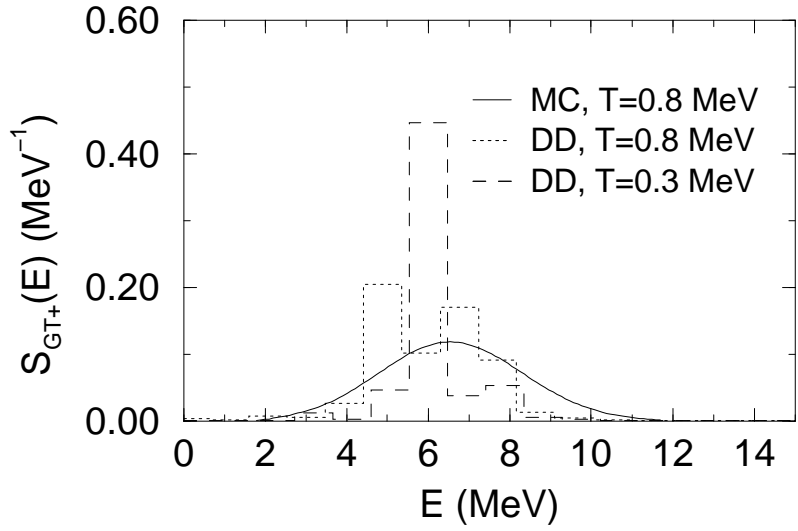
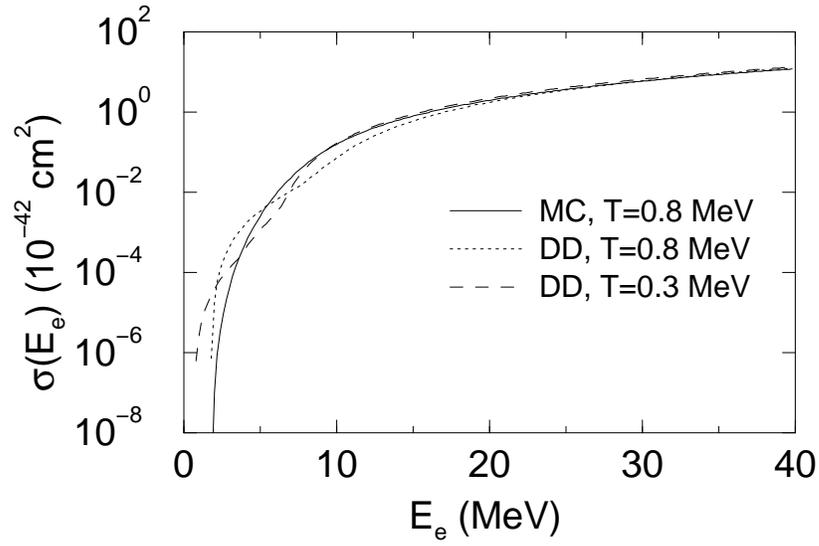
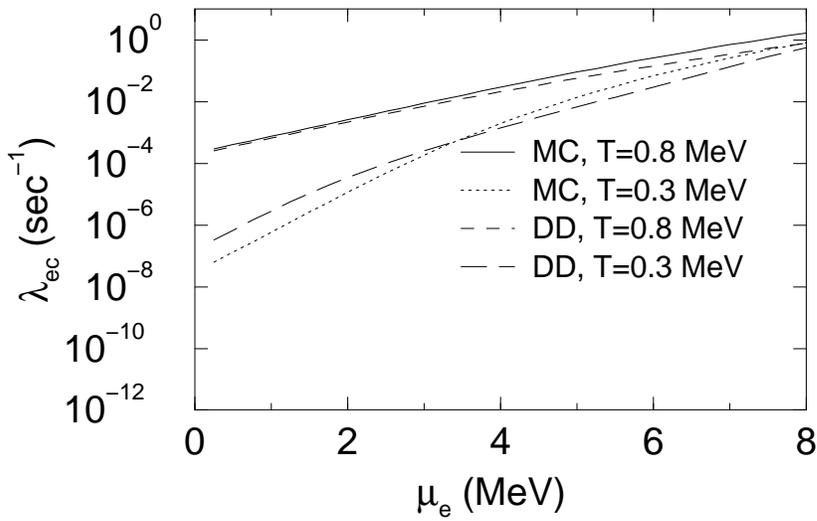
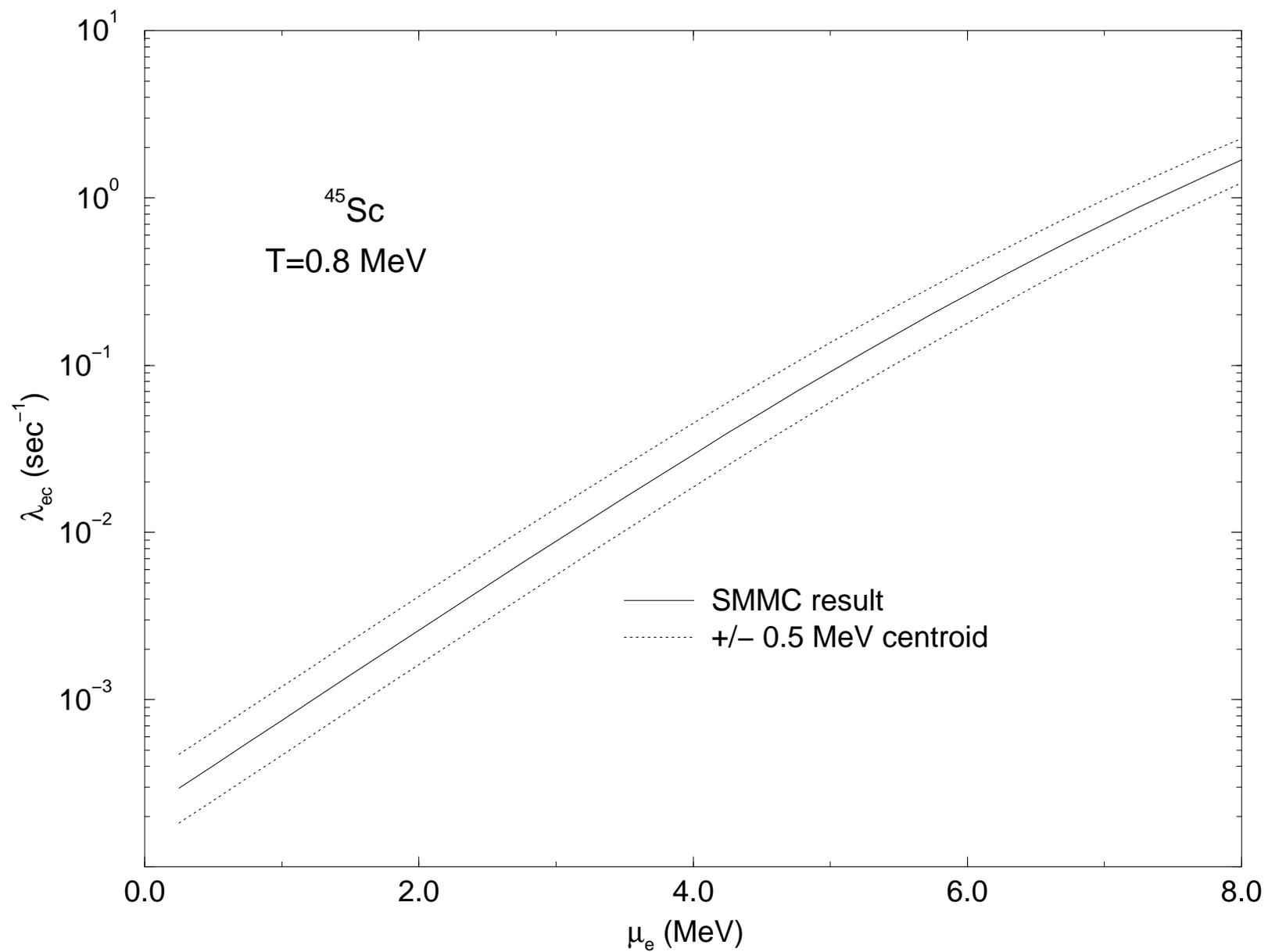
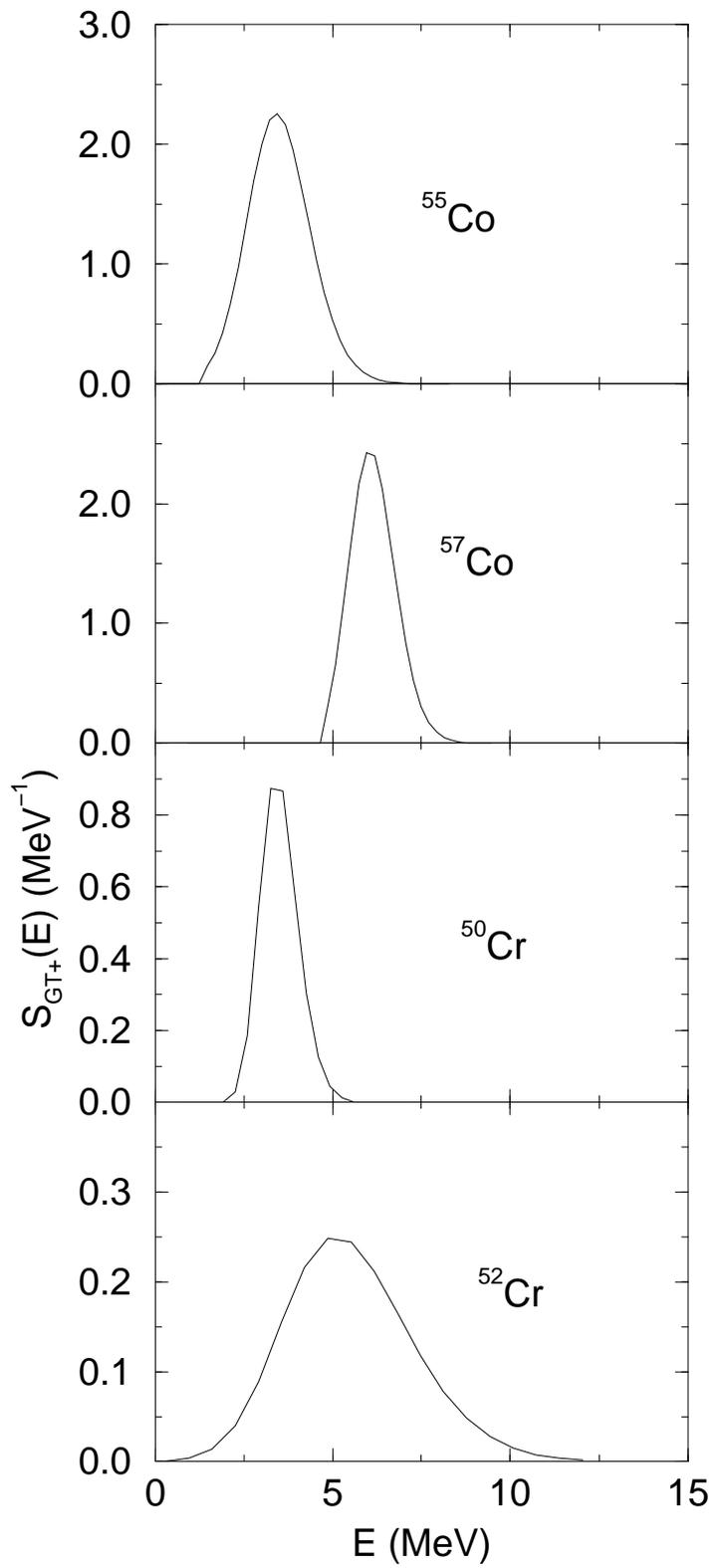
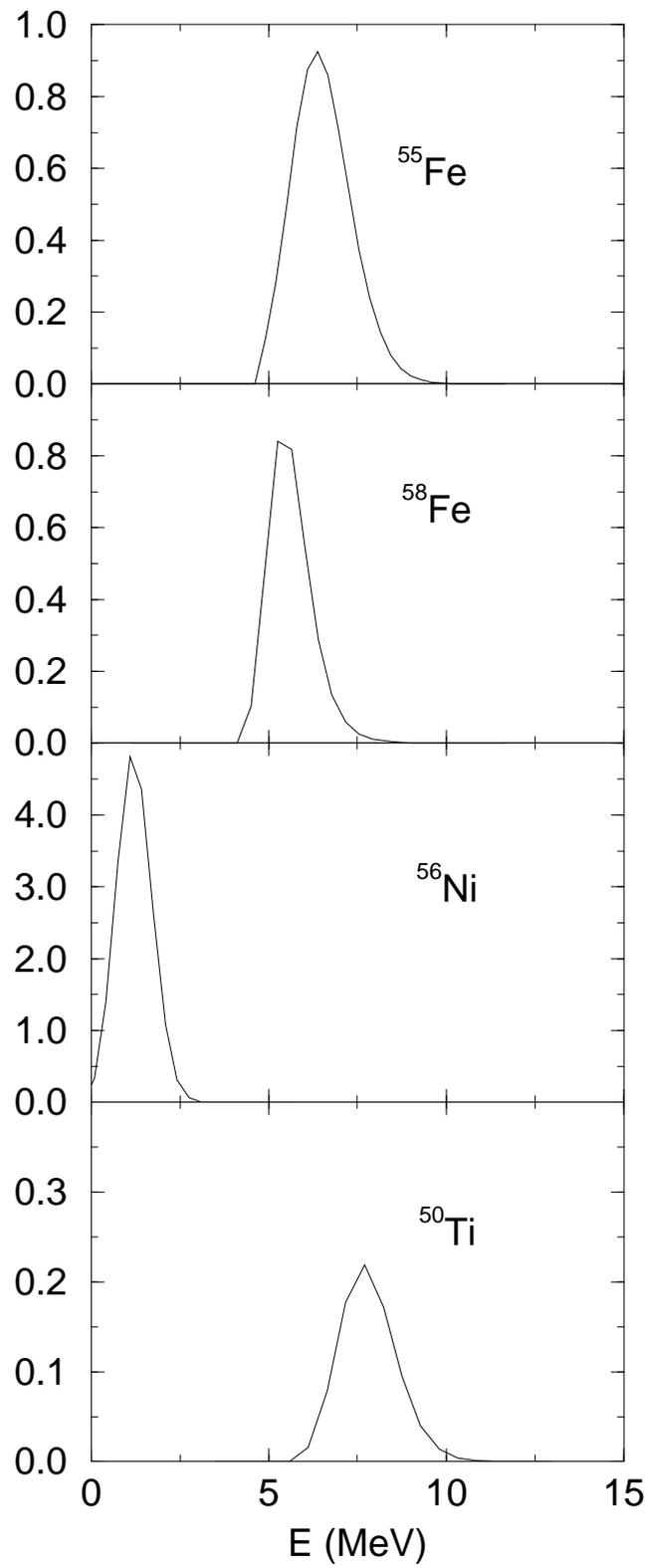
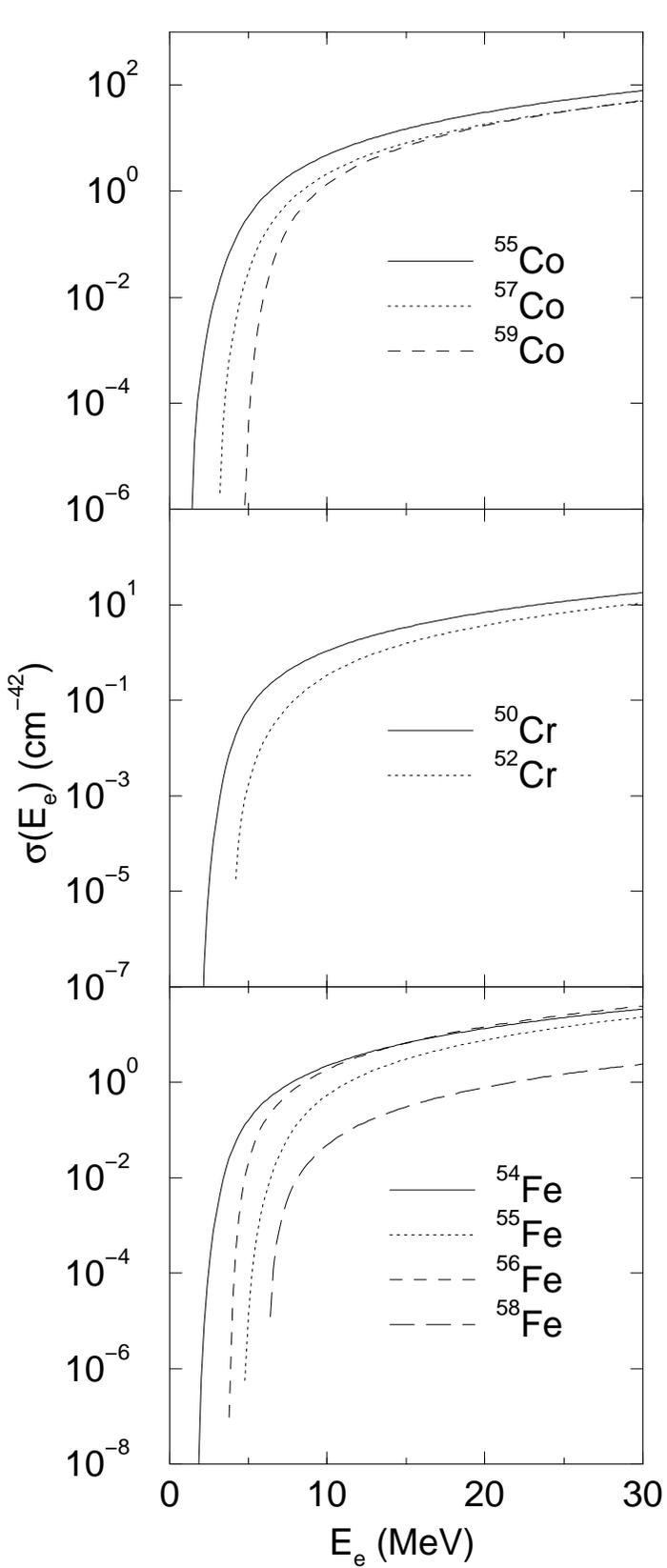
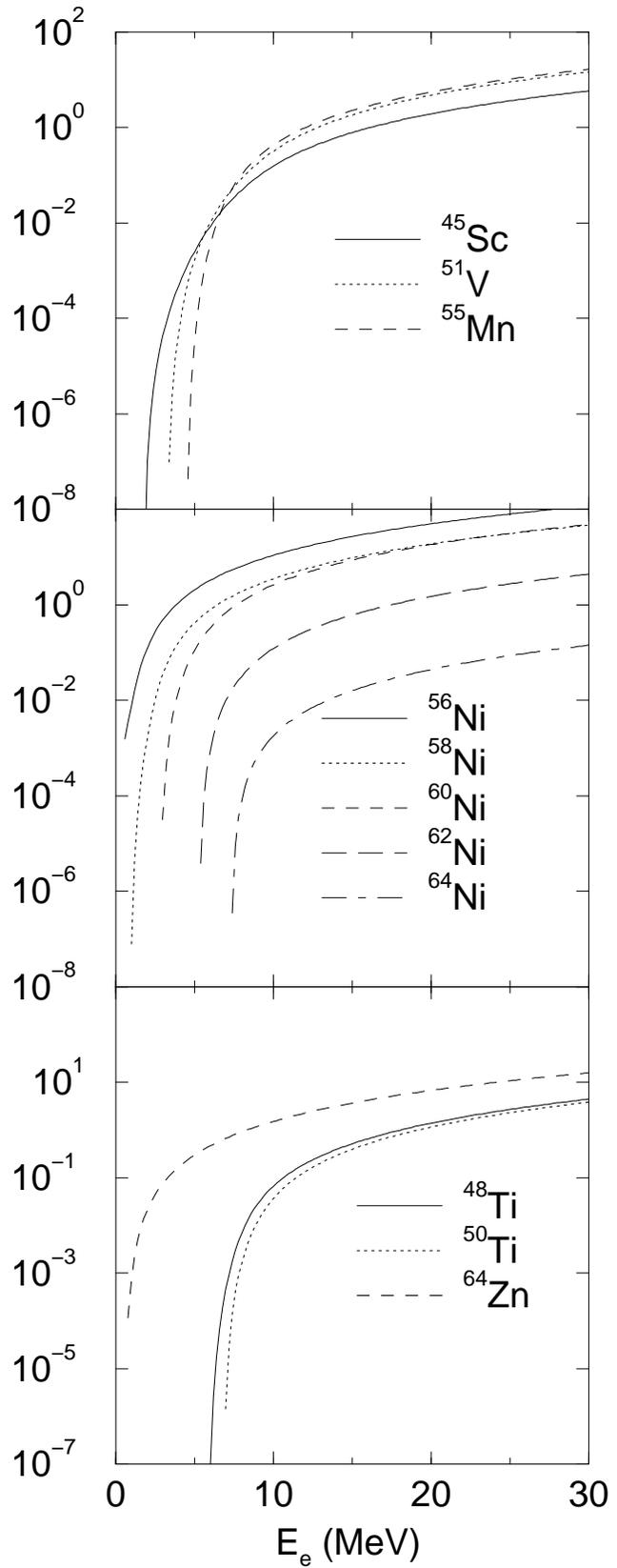
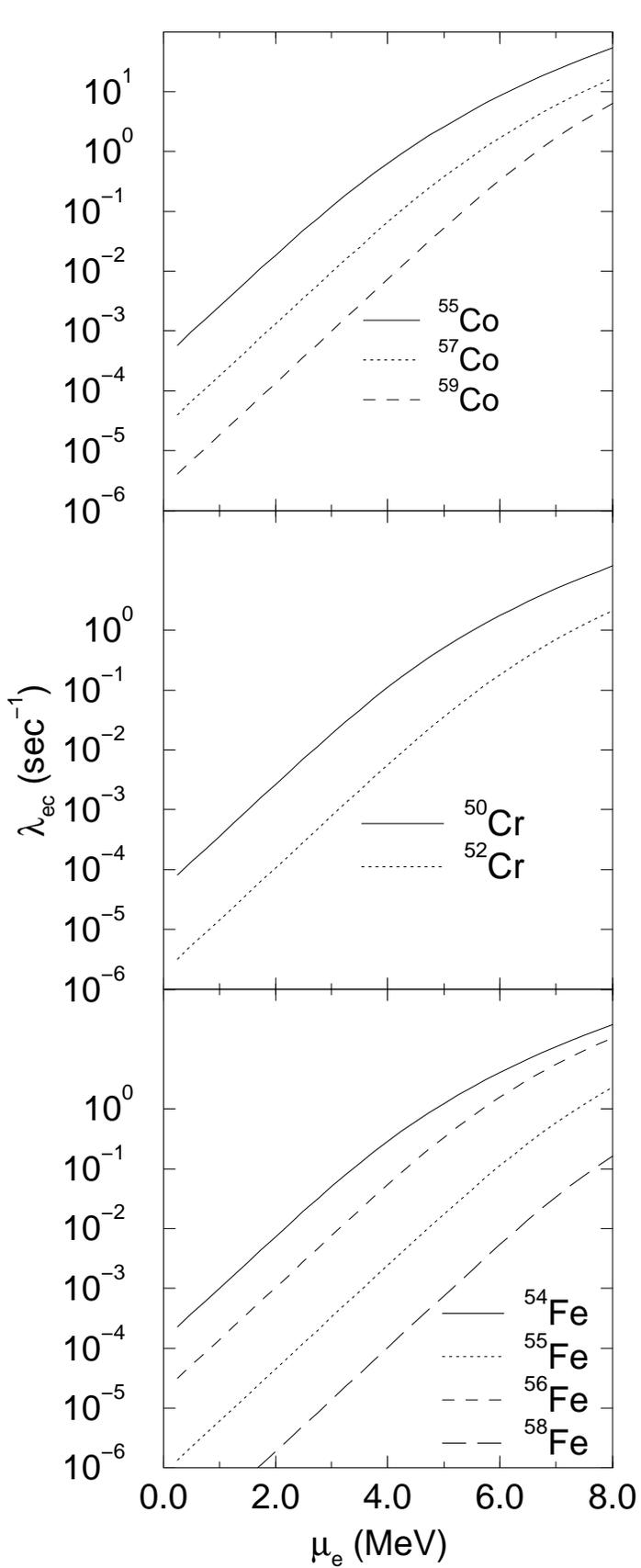
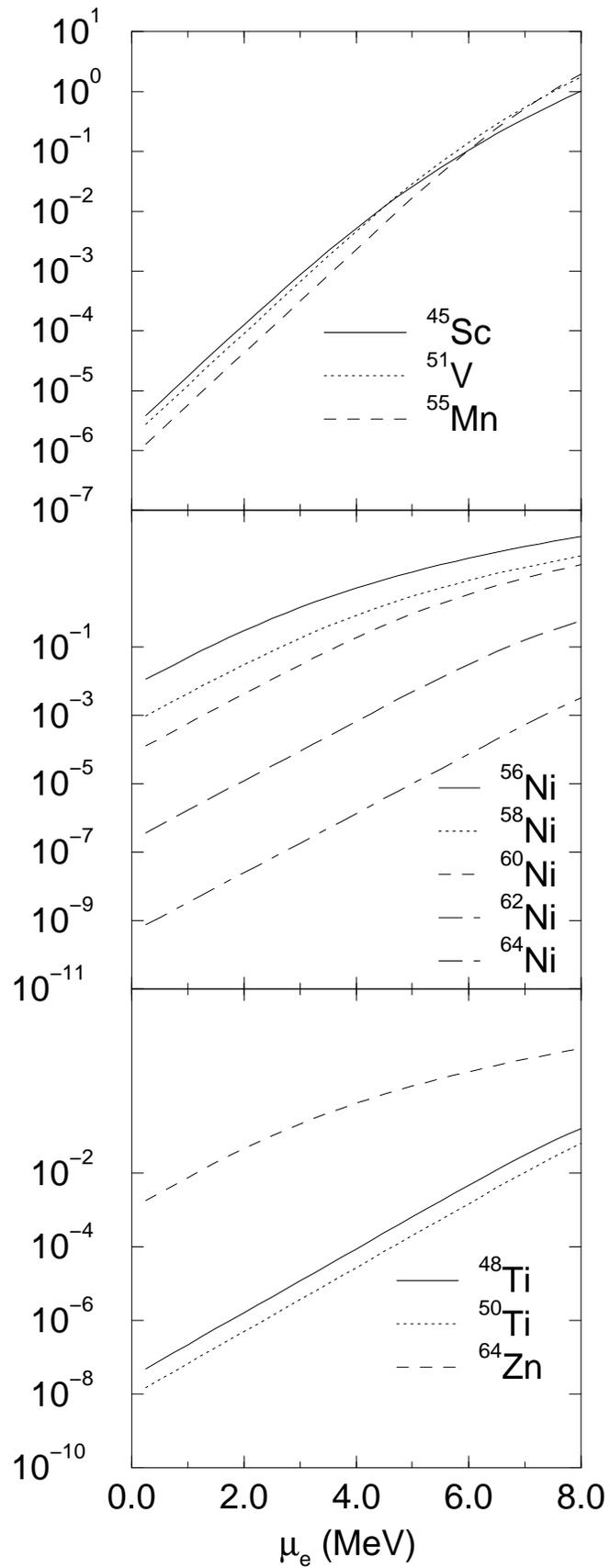
\end{document}